\documentclass{webofc}

\usepackage[varg]{txfonts}   
\usepackage{hyperref}
\usepackage{url}
\usepackage{lineno}
\usepackage{float}
\usepackage{placeins}
\RequirePackage{orcidlink}
\hypersetup{colorlinks=true,citecolor=blue,urlcolor=blue,linkcolor=blue}
\begin{document}
%
\title{Extending the reconstruction of displaced tracks at CMS during the HL-LHC with Line Segment Tracking}
%
%

\author{\firstname{Jade} \lastname{Chismar}\inst{1}\orcidlink{0009-0001-3662-6553}\fnsep\thanks{\email{jchismar@ucsd.edu}} on behalf of the CMS Collaboration\fnsep\thanks{Copyright 2026 CERN for the benefit of the CMS Collaboration. Reproduction of this article or parts of it is allowed as specified in the CC-BY-4.0 license.}
}

\institute{University of California, San Diego 
          }

\abstract{The upgrade of the Large Hadron Collider (LHC) to the High-Luminosity LHC (HL-LHC) will greatly increase the average number of simultaneous proton-proton interactions per bunch-crossing (pileup), and will place significant demands on computing resources for the reconstruction of charged-particle tracks. Line Segment Tracking (LST) is a novel, highly parallelizable algorithm that can run efficiently on GPUs and has been integrated into the CMS software to enable the reconstruction of displaced tracks in the high pileup scenarios of the HL-LHC. The LST algorithm is designed to build track candidates using only outer tracker hits, naturally allowing for high acceptance of displaced track signatures. In the HL-LHC tracking reconstruction sequence without LST, displaced tracking ends at a radial displacement of the track origin of approximately 8 cm. The usage of LST in track reconstruction enhanced efficiency for displaced tracks, extending the acceptance in radial displacement from the track origin by more than a factor of 4. In this work, we introduce a new LST object, the quadruplet (T4), defined as a linked pair of triplets with a common line segment, targeting displaced track signatures. We present the performance of this improved implementation of the LST algorithm in several physics scenarios, extending the displaced-track acceptance from 40 cm to 60 cm in radial displacement compared to LST without T4s.
}
\maketitle
\section{Introduction}
\label{intro}
With the upgrade of the Large Hadron Collider (LHC) to the High-Luminosity LHC (HL-LHC), new physics opportunities will come along with significant computational challenges for charged-particle track reconstruction. The increased luminosity is expected to result in high pileup (PU), with up to 200 simultaneous interactions per event, producing a large number of charged-particle tracks. This leads to increased processing time, necessitating algorithms that can keep up with the data rate. At the same time, displaced tracks, which are those that do not originate at the interaction point, represent one of the remaining phase spaces in which new physics may lie. Their reconstruction is particularly challenging in the high-occupancy environments expected at the HL-LHC, making it essential for tracking algorithms to efficiently reconstruct displaced tracks.

\begin{figure}[h!]
    \centering
    \includegraphics[width=\linewidth]{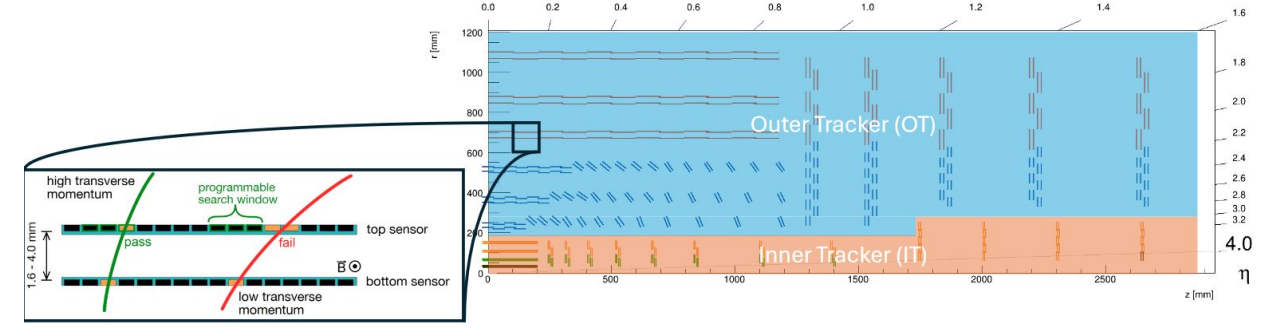}
    \caption{Sketch of the CMS Phase-2 Outer and Inner Trackers, with a detailed view of an Outer Tracker $p_\text{T}$ module.}
    \label{fig:ot}
\end{figure}

 
Line Segment Tracking (LST) is a parallelizable, hardware-agnostic algorithm designed for the CMS Phase-2 Outer Tracker (OT)~\cite{RefHLT, RefHLTold, Ref25-051}, shown in light blue in Figure~\ref{fig:ot}. The key characteristic of the CMS Phase-2 OT is transverse momentum ($p_\text{T}$) modules composed of 2 closely spaced silicon sensors~\cite{RefPhase2}. Linked pairs of hits in sensors of the same $p_\text{T}$ module represent minidoublets (MDs). The use of MDs reduces combinatorics, and allows for parallelization since MDs can be reconstructed independently using only local detector information.

\begin{figure}[h!]
    \centering
    \includegraphics[width=7cm,clip]{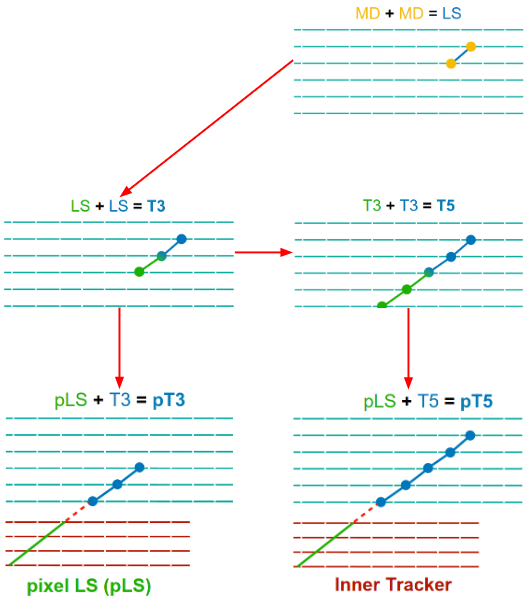}
    \caption{Sketch of the LST objects before this work. The horizontal red and green lines represent Inner Tracker and Outer Tracker layers, respectively. (Top Right): line segment as a pair of two MDs. (Middle Left): triplet (T3). (Middle Right): quintuplet. (Bottom Left): pixel seed with a triplet (pT3). (Bottom Right): pixel seed with a quintuplet (pT5).}
    \label{fig:lst}
\end{figure}

The LST algorithm starts from MDs and progressively links shorter objects to produce longer ones. First, pairs of MDs in neighboring layers are linked to create "line segments" (LSs). Linked pairs of LSs with a common MD are defined as triplets (T3s), and linked pairs of T3s with a common MD are defined as quintuplets (T5s)~\cite{RefML, RefHLT}. Finally, the LST algorithm uses a subset of Inner Tracker (IT) pixel seeds, referred to as pixel line segments (pLSs), to extend the reconstruction into the IT by linking them to LST objects. A pLS linked to a T5 forms a pixel quintuplet (pT5), while a pLS linked to a T3 forms a pixel triplet (pT3). The LST objects discussed so far are illustrated in Figure~\ref{fig:lst}. 

A final list of LST Track Candidates (TCs) is created by collecting pT5s, pT3s, T5s, and unused pLSs with at least 4 hits. The pT5s objects are the main tracking efficiency driver of the LST algorithm, and the pT3s are used for further efficiency recovery. The T5s, as OT-only objects, drive the efficiency of displaced tracks, while pLSs dominate the efficiency of tracks at high pseudorapidity ($\eta$) and low $p_\text{T}$.

Objects of a given type can be constructed independently of each other, which leads to massive parallelization. The usage of OT hits allows for the acceptance of displaced track signatures. Machine learning is utilized in the creation of more complicated, longer objects. Specifically, shallow deep neural networks (DNNs) are used for object selection, with no effect on timing, while leading to a significant reduction in fake rate and track duplicates~\cite{RefML, RefMLNew}.

\section{Quadruplet Object}
\label{sec-T4}
\begin{figure}[h!]
\centering
    \includegraphics[width=0.5\linewidth]{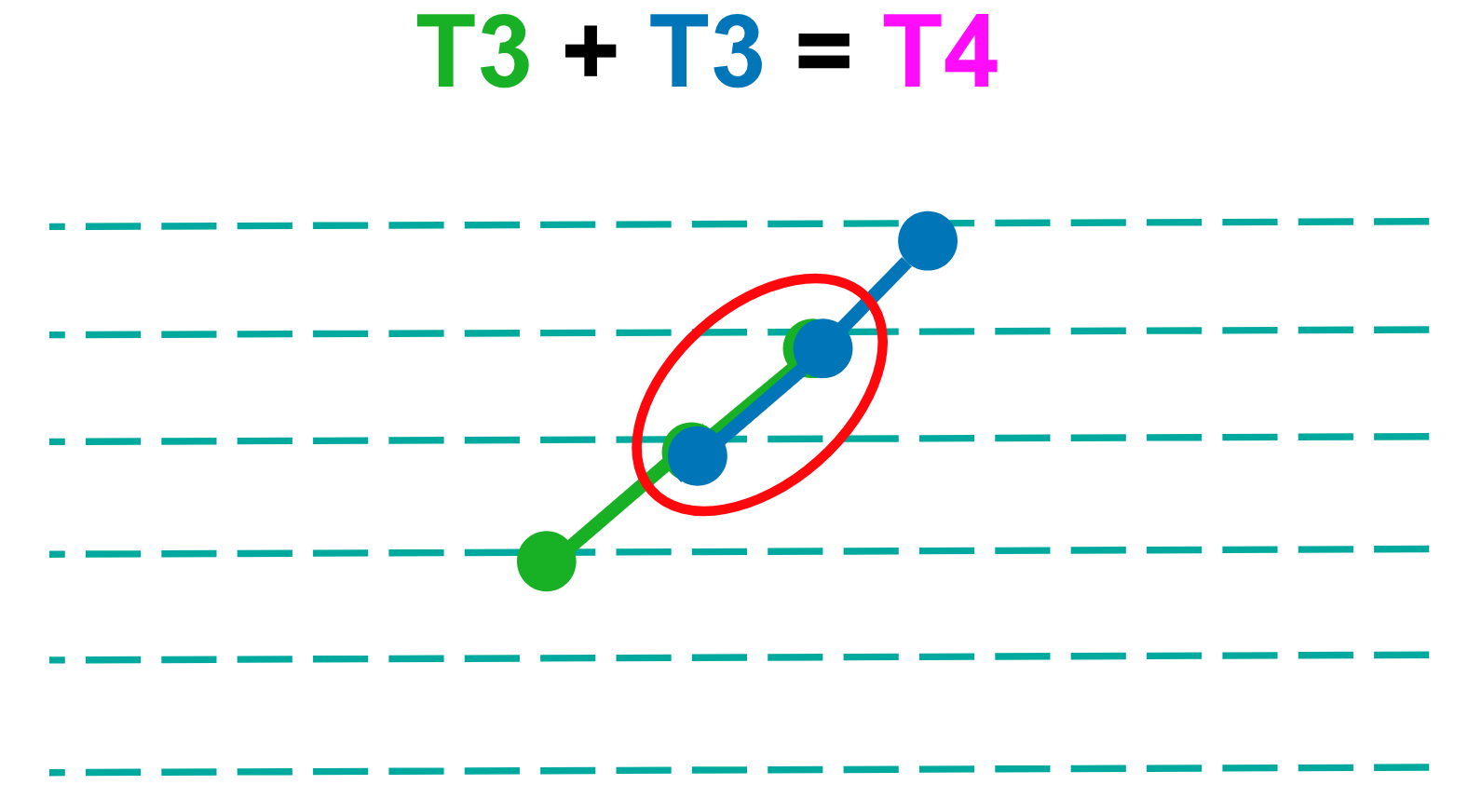} 
    \caption{Sketch of the quadruplet (T4) object: a linked pair of T3s with a shared LS.}
    \label{fig:T4}
\end{figure}

Quadruplets (T4s) are a new LST object type targeting displaced tracks. They are defined as a linked pair of T3s with a common LS, as shown in Figure~\ref{fig:T4}. 
T4s are only built using T3s that are unused in pT5s, pT3s, or T5s, and in regions that are not covered by T5s: in the full endcaps and in the third layer of the OT. 

Machine learning is used to reduce fake and duplicate rates via a multi-class DNN classifier. Because prompt tracks vastly outnumber displaced tracks, a three-class classifier is used to distinguish displaced tracks from the more abundant prompt and fake tracks. The DNN classifies T4 candidates as real prompt, real displaced, or fake. Displaced candidates are defined as simulated tracks originating more than 10 cm from the primary interaction point in the transverse plane. The input features of the DNN describe the hit coordinates and circle geometry, including the difference in $\eta$, $\phi$, r, and z coordinates of successive hits, as well as the T3 DNN output scores of the constituent T3s. The DNN selection thresholds, tuned to achieve a true displaced T4 retention ranging from 95\% to 99\%, and binned by $p_T$ and $\eta$, are set to accept real displaced candidates while rejecting real prompt and fake ones. Additional requirements are imposed for track consistency in $\mathrm{r-}\varphi$ and $\mathrm{r-}\mathrm{z}$ planes. An explicit cut on the lower bound on the transverse impact parameter ($\mathrm{d}_0$) is also applied to further decrease the fake rate from prompt tracks. The T4 candidates that pass the selections are added as TCs, resulting in the new final TC collection consisting of pT5s, pT3s, T5s, T4s, and pLSs.

\section{Physics Performance}
The metrics used for physics performance assessment are the tracking efficiency and fake rate. Efficiency is defined as the fraction of simulated tracks matched to at least one reconstructed track, while the fake rate represents the fraction of misidentified reconstructed tracks. A simulated track is considered matched to a reconstructed one if more than 75\% of the hits of the reconstructed track originate from the same simulated track.

The tested configurations are offline tracking ones, restricted to the first two tracking iterations. In the legacy tracking configuration, the track seeding step is performed in two iterations by the legacy pixel tracking algorithm~\cite{reco}, creating seeds with 3 or 4 hits. The track building step is also performed in two iterations, using the Combinatorial Kalman Filter (CKF) algorithm~\cite{kalman}. In the LST tracking configuration, the legacy pixel tracking seeds produced as in the legacy configuration are combined with the OT objects with $p_T > 0.8$ GeV by the LST algorithm within the OT acceptance of $|\eta| < 2.5$.
No additional building step is performed after the LST, and therefore no pLS building is carried out. Pixel tracks with fewer than four hits are discarded. The high-purity selections are relaxed. In both configurations, the track fitting step is performed with the CKF algorithm.

The tracking performance of LST as a track building algorithm~\cite{reco} is first evaluated on a displaced-muon sample generated with a muon gun, without PU. The sample contains 10 muons per event, with their production vertices uniformly distributed across a 1 $\mathrm{m}^3$ cube ($\pm$ 50 cm from the origin). 

\begin{figure}[h!]
\centering
    \includegraphics[width=0.6\linewidth]{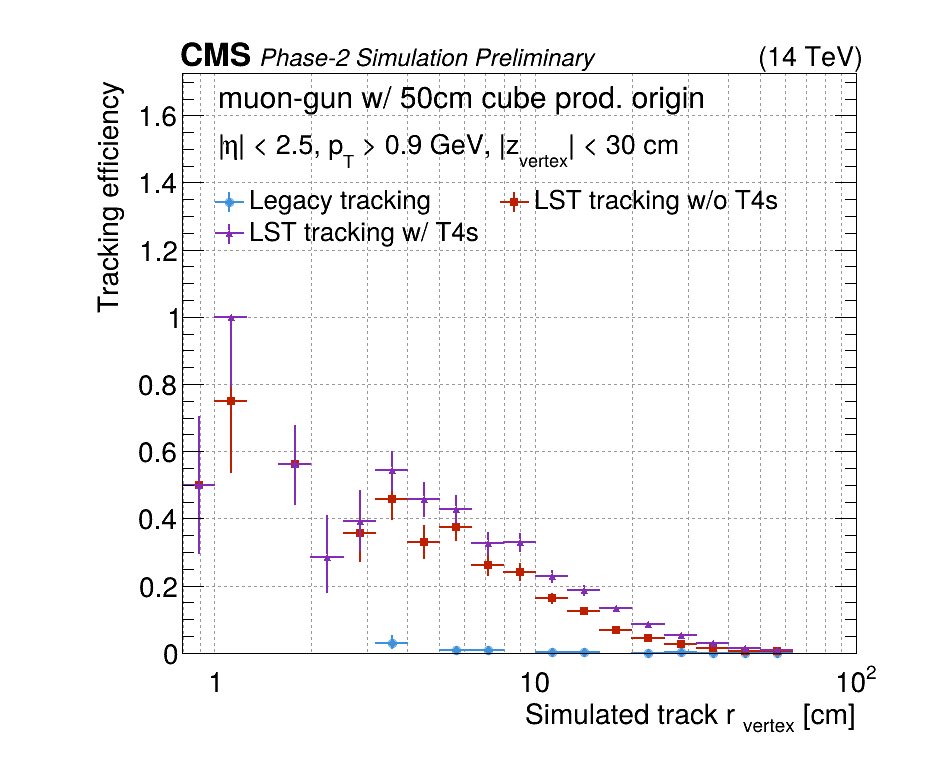}
    \caption{The tracking efficiency is shown as a function of the radial displacement of the simulated track vertex ($\mathrm{r}_\mathrm{vertex}$) for legacy tracking (blue circles), LST tracking without T4s (red squares), and LST tracking with T4s (purple triangles), for simulated tracks with $p_\text{T}>0.9$ GeV, $|\eta|<2.5$, and $|\mathrm{z}_\mathrm{vertex}|<30$ cm for 10,000 muon-gun events. The empty bins correspond to bins with zero entries in either the numerator or the denominator.~\cite{T4DP}}
    \label{fig:muon-eff}
\end{figure}

The tracking efficiency as a function of the radial displacement of the simulated track vertex ($\mathrm{r}_\mathrm{vertex}$) is shown in Figure~\ref{fig:muon-eff}. The addition of OT tracks by the LST algorithm greatly increases the efficiency and extends the acceptance of displaced tracks up to $\sim$40 cm compared to the legacy configuration. The addition of T4s to LST shows a significant increase in displaced track efficiency, with tracking efficiency doubling starting from $\mathrm{r}_\mathrm{vertex} \sim$15 cm.

\begin{figure}[h!]
\centering
    \includegraphics[width=0.6\linewidth]{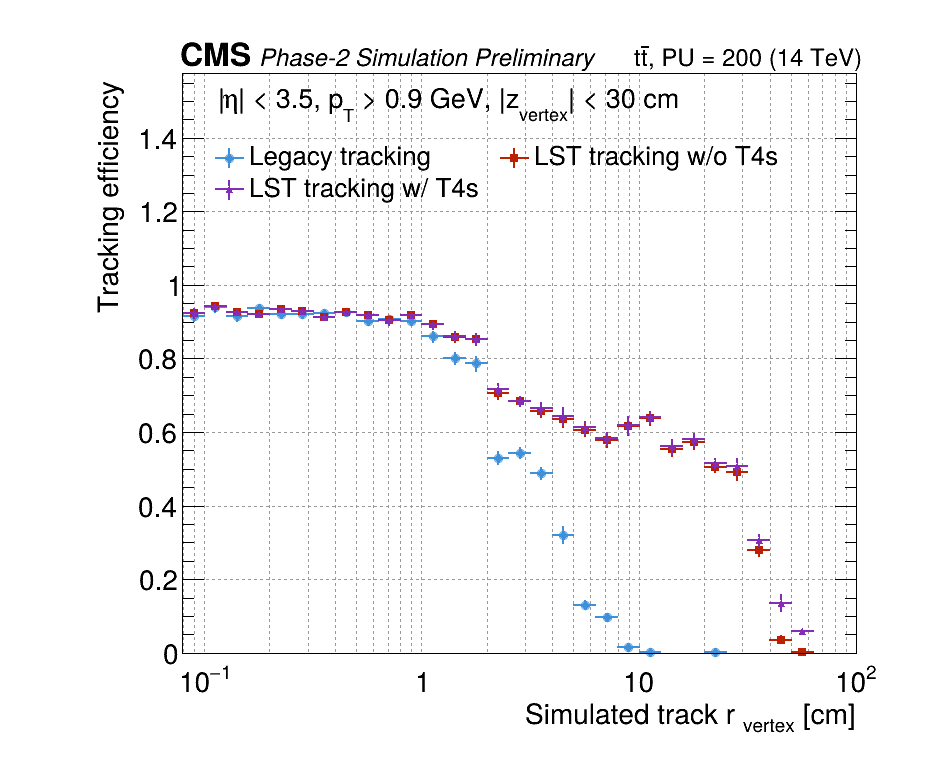}
    \caption{The tracking efficiency is shown as a function of the radial displacement of the simulated track vertex ($r_{\mathrm{vertex}}$) for legacy tracking (blue circles), LST tracking without T4s (red squares), and LST tracking with T4s (purple triangles) for 1000 $t\bar t$ events at PU=200. The empty bins correspond to bins with zero entries in either the numerator or the denominator.~\cite{T4DP}}
    \label{fig:ttbar-eff-r}
\end{figure}

\begin{figure}[h]
    \centering
    \includegraphics[width=\linewidth]{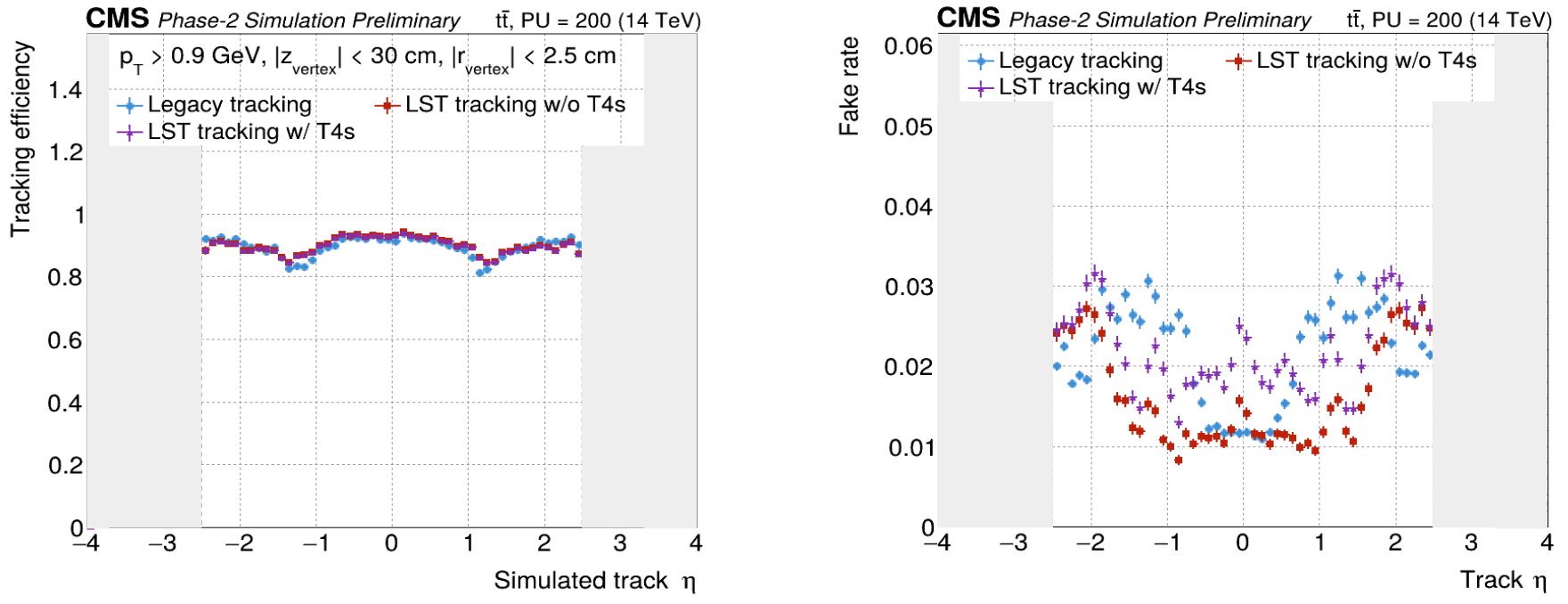}
    \caption{The tracking efficiency (left) and fake rate (right) are shown as a function of the simulated track pseudorapidity and reconstructed track pseudorapidity ($\eta$), respectively, for legacy tracking (blue circles), LST tracking without T4s (red squares), and LST tracking with T4s (purple triangles) for 1000 $t\bar t$ events at PU=200. The shading at high $\eta$ highlights the region beyond the OT coverage, which defines the LST acceptance ($|\eta|<2.5$).~\cite{T4DP}}
    \label{fig:dup-fake}
\end{figure}

The tracking performance is then evaluated in $t\bar t$ events at PU=200, providing a more comprehensive assessment of LST in a realistic high-pileup environment, as shown in Figures~\ref{fig:ttbar-eff-r} and~\ref{fig:dup-fake}. The LST algorithm extends the displaced track acceptance from approximately 10 cm to about 40 cm, as compared to the legacy tracking, while prompt efficiency stays at the same level. The addition of T4s to LST further extends displaced track acceptance to an $\mathrm{r}_\mathrm{vertex}$ of 60 cm. There are minimal changes in prompt efficiency with the addition of T4s for OT acceptance of $|\eta|<2.5$ (left), and minor increases in fake and duplicate rates (right), shown in  Figure~\ref{fig:dup-fake}.
\section{Summary}
The LST algorithm enables the reconstruction of displaced tracks in CMS at the HL-LHC, primarily relying on OT-only objects. The newly introduced LST T4 objects further extend the acceptance for displaced tracks, with only minor increases in the fake and duplicate rates. T4 objects are now part of the LST algorithm in the CMS software, enabling their use in the tracking reconstruction at the HL-LHC.

\vspace{0.5cm}
\noindent
\small{This work was supported by the National Science Foundation under Cooperative Agreements OAC1836650 and PHY-2323298.}

%
\FloatBarrier


\begin{thebibliography}{}
%
%
\bibitem{RefHLTold}
CMS Collaboration, Performance of the Line Segment Tracking Algorithm in the CMS Phase-2 High Level Trigger Tracking. CMS Detector Performance Summary CMS-DP-2024-014 (2024). \url{https://cds.cern.ch/record/2890677}

\bibitem{Ref25-051}
CMS Collaboration, Performance of mkFit and LST algorithms in the CMS Phase-2 High Level Trigger Tracking. CMS Detector Performance Summary CMS-DP-2025-051 (2025). \url{https://cds.cern.ch/record/2941438}

\bibitem{RefHLT}
CMS Collaboration, A heterogeneous \& vectorized sequence for the Phase-2 HLT tracking reconstruction. CMS Detector Performance Summary CMS-DP-2026-026 (2026). \url{https://cds.cern.ch/record/2961608}

\bibitem{RefPhase2}
CMS Collaboration, The Phase-2 Upgrade of the CMS Tracker. CERN-LHCC-2017-009, CMS-TDR-014 (2017). \url{https://cds.cern.ch/record/2272264}

\bibitem{RefML}
CMS Collaboration, Improved Performance of Line Segment Tracking Using Machine Learning. CMS Detector Performance Summary CMS-DP-2023-075 (2023). \url{https://cds.cern.ch/record/2872904}

\bibitem{RefMLNew}
CMS Collaboration, Extended ML Selections and Track Embeddings for Duplicate Removal in Line Segment Tracking (LST). CMS Detector Performance Summary CMS-DP-2025-048 (2025). \url{https://cds.cern.ch/record/2941435}

\bibitem{reco}
CMS Collaboration, Description and performance of track and primary-vertex reconstruction with the CMS tracker. JINST 9 P10009 (2014). \href{https://doi.org/10.1088/1748-0221/9/10/P10009}{DOI:10.1088/1748-0221/9/10/P10009}, \url{https://arxiv.org/abs/1405.6569}

\bibitem{kalman}
R. Frühwirth, Application of Kalman filtering to track and vertex fitting. Nucl. Instrum. Meth. A 262 444 (1987). \href{https://doi.org/10.1016/0168-9002(87)90887-4}{DOI:10.1016/0168-9002(87)90887-4}

\bibitem{T4DP}
CMS Collaboration, Extending the reconstruction of Phase-2 displaced tracks using Line Segment Tracking. CMS Detector Performance Summary CMS-DP-2026-029 (2026). \url{https://cds.cern.ch/record/2961685}
\end{thebibliography}
\end{document}